\documentclass[a4paper,11pt]{article}
\pdfoutput=1 % if your are submitting a pdflatex (i.e. if you have
\usepackage{jcappub} % for details on the use of the package, please
\usepackage{gensymb}
\usepackage[]{algorithm2e}% added by jorit
\usepackage{subcaption}%added by jorit
\usepackage{sidecap}%added by jorit
\usepackage{aasmacros}%added by jorit

\usepackage[T1]{fontenc} % if needed

\newcommand{\abcq}{{{qABC}}}

\defcitealias{herbel2017redshift}{H17}

\usepackage{array}
\newcolumntype{L}[1]{>{\raggedright\let\newline\\\arraybackslash\hspace{0pt}}m{#1}}
\newcolumntype{C}[1]{>{\centering\let\newline\\\arraybackslash\hspace{0pt}}m{#1}}
\newcolumntype{R}[1]{>{\raggedleft\let\newline\\\arraybackslash\hspace{0pt}}m{#1}}
\title{\boldmath Accelerating Approximate Bayesian Computation with Quantile Regression: Application to Cosmological Redshift Distributions}

 \author[]{T. Kacprzak,}
 \author[]{J. Herbel,}
 \author[]{A. Amara,}
 \author[]{and A. R\'{e}fr\'{e}gier}
 \affiliation[]{Department of Physics, Eidgen\"ossische Technische Hochschule Z\"urich,\\Wolfgang-Pauli-Str. 27, 8093 Z\"{u}rich, Switzerland}

\emailAdd{tomasz.kacprzak@phys.ethz.ch}
\notoc

\abstract{
Approximate Bayesian Computation (ABC) is a method to obtain a posterior distribution without a likelihood function, using simulations and a set of distance metrics.
For that reason, it has recently been gaining popularity as an analysis tool in cosmology and astrophysics.
Its drawback, however, is a slow convergence rate.
We propose a novel method, which we call \abcq, to accelerate ABC with Quantile Regression.
In this method, we create a model of quantiles of distance measure as a function of input parameters.
This model is trained on a small number of simulations and estimates which regions of the prior space are likely to be accepted into the posterior.
Other regions are then immediately rejected.
This procedure is then repeated as more simulations are available.
We apply it to the practical problem of estimation of redshift distribution of cosmological samples, using forward modelling developed in previous work.
The \abcq\ method converges to nearly same posterior as the basic ABC.
It uses, however, only 20\% of the number of simulations compared to basic ABC, achieving a fivefold gain in execution time for our problem.
For other problems the acceleration rate may vary; it depends on how close the prior is to the final posterior.
We discuss possible improvements and extensions to this method.
}

\begin{document}
\maketitle
\flushbottom

\newpage

\section{Introduction}

For many inference problems, it is impossible or impractical to create a likelihood function of data given model parameters.
Approximate Bayesian Computation (ABC) is a method to approximate the posterior distribution without using a likelihood function \citep{marin2012approximate,csillery2010approximate}.
This is achieved by generating simulations from a model and comparing them to the data using distance metrics.
Due to this property, ABC has recently been gaining popularity in cosmology and astrophysics.
It has been used for constraining cosmological parameters  \citep{weyant2013likelihood,jennings2016approach,jennings2016astroabc}, studying substructure content of strong gravitational lenses \citep{birrer2017lensing}, estimating parameters of galaxy evolution models \citep{carassou2017inferring}, and measuring distributions of shapes and sizes of observed galaxies \citep{akeret2015approximate}.
Recently \citep[][hereafter \citetalias{herbel2017redshift}]{herbel2017redshift} presented a way of measuring the redshift distribution of cosmological samples with ABC.

In the most basic ABC formulation, a sample from the prior is accepted into the posterior if their distance metric is lower than some chosen threshold.
If the chosen threshold is too low, the acceptance rate will be low, and many simulator runs will be required.
This will result in a slow convergence of the algorithm.
If the threshold is high, then the approximated posterior will be much broader than the true posterior.
In the extreme case of a very high threshold, all prior samples will be accepted and no information will be gained.

For many problems in practice, running a single simulation can be computationally expensive.
It is therefore important to reduce the number of simulations to a minimum.
The basic ABC algorithm typically requires a large number of simulations, which can significantly limit its practical usability.
Several algorithms have been proposed for accelerating the ABC method.
It has been implemented within the Monte Carlo Markov Chain (MCMC) framework \citep{majoram2003markov,marin2012approximate}, using Sequential Monte Carlo \citep{sisson2007sequential,toni2009simulation} and Population Monte Carlo \citep{akeret2015approximate}.
While these methods explore the parameter space more efficiently than the basic ABC algorithm, they use the available information in a limited way:
due to the Markov property, the choice of a new point is only informed by the previous one, and thus the earlier simulations are ``forgotten''.
Since every simulation sample often comes at a high computational cost and therefore constitutes a precious piece of information, an efficient method would aim to utilise all available simulations when accepting or rejecting samples.

The result of ABC depends on the conditional distribution $p(d|\theta)$ of distance measure $d$ given model parameters $\theta$.
Both basic ABC and its Monte Carlo implementations do not make strong assumptions about the shape of this distribution.
The basic ABC method will work accurately for any $p(d|\theta)$, regardless of its properties.
The MCMC-enabled methods are also robust and yield accurate results.
The price for that consistency is the slow convergence.

However, for many ABC applications, a valid assumption can be made that $p(d|\theta)$ varies smoothly in the $\theta$ parameter space.
A model for a smooth $p(d|\theta)$ can then be created an trained on prior samples, for which the simulations and distance measures were already computed.
This model can then make a prediction of $p(d|\theta)$ for new points in prior parameter space $\theta$.
Such a model can then be used for accelerating the convergence of ABC.

One way to use this model is through Sequential History Matching \citep{craig1997pressure}.
This approach quickly excludes those parts of prior space, for which, according to the model, no samples are expected to be accepted to the posterior.
These regions are marked as infeasible and removed from further analysis.
The exclusion steps are performed iteratively as the models are refined with more training data.
In \citep{wilkinson2014accelerating}, a Gaussian Process (GP) method has been used to model the $p(d|\theta)$ distribution.
In this case, the decision about which prior samples to reject is made using the uncertainty on the mean of distance measure, as calculated by the GP.
This method has been demonstrated to perform well on the trial data sets.

In the GP model presented in \citep{wilkinson2014accelerating}, the distribution $p(d|\theta)$ is assumed to be Gaussian.
While for a general GP the noise distribution does not have to be Gaussian (implementations with common noise models exist, such as Student-T, Poisson, Heteroscedastic Gaussian, warped-GP \citep{vanhatalo2009gaussian,snelson04warpedgaussian,lequoc2005heteroscedastic,vanhatalo2012bayesian}), the choice of distribution has to be made a priori, as it is a part of the model.
This assumption may be too restrictive for some practical applications.
One reason for this is that $p(d|\theta)$ often does not have a global closed form distribution, but can instead change its properties across the $\theta$ parameter space.
In that case GP will struggle to model the space efficiently.
Moreover, if several distance measures are used simultaneously, that choice would have to be made for each distance measure separately.

To address these problems, we present a new approach to model the distribution $p(d|\theta)$, based on Quantile Regression (QR) \citep{koenker78regressionquantiles,takeuchi06nonparametricquantile,yu2003quantile}, which we call \abcq.
Just as regression with least squares finds the mean of the data as a function of input parameters, quantile regression finds the value corresponding to a chosen quantile.
In the \abcq\ method, for any value of input parameter space $\theta$, we find the value of distance metric $d_q$, which corresponds to a quantile $q$, such that the cumulative probability $P(d_q|\theta) = q$.
This function $d_q(\theta)$ is assumed to vary smoothly with $\theta$.
When $q$ is chosen to be small, we effectively model the ``bottom'' of the distance measure distribution.
ABC algorithms are sensitive to exactly this part of the $p(d|\theta)$ distribution, as thresholds are usually chosen to be as low as possible.
This method can easily handle complicated $p(d|\theta)$ distributions, as long as the smoothness assumption is fulfilled.
In particular, it is well suited for the common practical case  where the distance measure is positive definite.

The prior on the function space of the quantile model $d_q(\theta)$ is created in reproducing kernel Hilbert space (RKHS).
It is a non-parametric model, which does not assume any functional form for the quantile function itself; instead it controls its smoothness using a positive definite kernel \citep{hofman2008kernel}.
We use a Support Vector Machine \citep{cristianini2000introduction} implementation of quantile regression \citep{hwang2005simplequantile,crambes2013support,steinwart2011estimating,steinwart2017liquid}.
This method is also well suited for high dimensional input parameter spaces.

Once the quantile model is created, we use a variant of sequential history matching based on the procedure in \citep{wilkinson2014accelerating}.
We use the model to reject infeasible parts of the $\theta$ parameter space; new simulations will only be ran for the points in regions, which were not excluded.
When more simulations are available, the model is re-calculated and the next rejection step is performed.
This process can continue until the ABC algorithm reaches the desired level of convergence.

As an application of the \abcq\ method we consider the measurement of the distribution of redshifts of cosmological galaxy samples, using a dataset from \citetalias{herbel2017redshift}.
They measured the distribution of redshifts $n(z)$ of population of galaxies in the COSMOS\footnote{\url{http://cosmos.astro.caltech.edu/}} field.
They used imaging data in that field, obtained by an optical telescope, as well as a spectroscopy data from VVDS survey\footnote{\url{ http://cesam.oamp.fr/vvdsproject/vvds.htm}} area.
A forward modelling approach was used to simulate images and spectra.
The model had 31 input parameters that controlled the redshift dependence of galaxy magnitudes, colours, and sizes.
The redshifts of the galaxies detected in the simulated images yielded an $n(z)$ distribution corresponding to the input parameters.

A basic ABC algorithm was used in that work.
The number of samples from the prior, for which the simulation was evaluated, was 140000.
Running this number of simulations required significant amount of computing time.
Here, we use the full dataset from \citetalias{herbel2017redshift} and investigate if the \abcq\ method can achieve faster convergence.

This paper is organised as follows.
In Section \ref{sec:method} we demonstrate the \abcq\ method on a toy model example.
The description of application to the redshift distribution measurement problem is described in Section \ref{sec:application}.
In Section \ref{sec:results} we present the results. We conclude in Section \ref{sec:conclusions}, as well as discuss future prospects and possible extensions of this method.

\section{The \abcq\ method}
\label{sec:method}

The \abcq\ method aims to create a model of quantile function $d_q$ of the conditional distribution $p(d|\theta)$ of distance $d$ given a set of parameters $\theta$, for a set of quantiles $q$.
This model is trained with all available simulations.
It is then used to make a prediction of $d_q$ for each prior sample.
In the next step, we determine which samples are \emph{infeasible}: that are very unlikely to be accepted into the posterior, according to the quantile model.
If the quantile function $d_q$ of a sample is large compared to other samples, this sample is deemed infeasible and rejected from further analysis.
This way the prior volume shrinks and fewer simulations are needed.

A good degree of confidence is therefore needed before discarding an estimated infeasible region.
In the presence of limited training data, however, the predicted value of quantile function will be uncertain.
This uncertainty depends on the number of training samples: the more training samples are used, the lower the uncertainty on the quantile function.
Quantile regression algorithm does not natively provide uncertainty estimates on predicted quantiles.
To address this, we estimate the uncertainty $\sigma[d_q]$ using a simple resampling approach, which was inspired by the jacknife method.
In our approach, we estimate the quantile function $N$ times, each time leaving out a small fraction of the data.
The resulting $N$ functions are then used to estimate the uncertainty $\sigma[d_q]$ on $d_q(\theta)$.
We use a median of $N$ functions as the central value, and median absolute deviation as the error estimate.
We next describe the details of the application of the method using a toy model.

\begin{figure}
\includegraphics[width=1.0\textwidth]{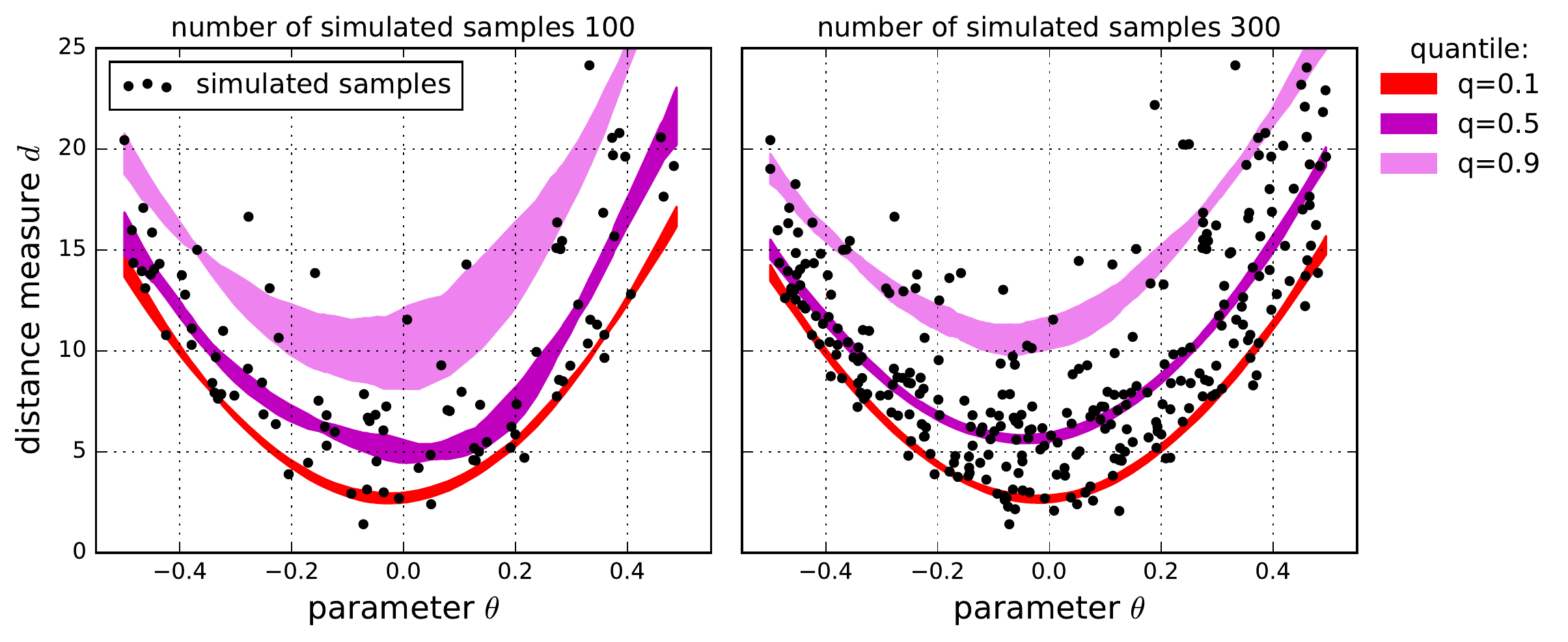}
\caption{Demonstration of quantile regression.
The black points correspond to the simulated data, and the colourful bands to predicted quantile functions.
The width of the band corresponds to the uncertainty on the quantile function.
The uncertainty was estimated using the method described in Section \ref{sec:method}.
The quantile uncertainty decreases when more training data is used.
}
\label{fig:demo1}
\end{figure}

\subsection{Toy model}
\label{sec:toy}

We create a simple toy model to demonstrate the \abcq\ method.
First, we create an example data set with one parameter $\theta \in (-1, 1)$ and single distance measure $d$.
This distance measure $d$ depends on $\theta$ with noise drawn from $\chi^2$ distribution with 5 degrees of freedom: $d \sim  1 + 50  \theta^2 + | 1+\theta | \chi^2_5$.
The shape of $p(d|\theta)$ therefore also depends on $\theta$.
% In this simple scenario, the random variable does not depend on $\theta$, but this is not a requirement of our method.
We use the \textsc{LiquidSVM}\footnote{\url{http://www.isa.uni-stuttgart.de/software/}} package \citep{steinwart2017liquid} to perform quantile regression.
This algorithm automatically performs the kernel parameter selection using integrated cross-validation.
Configuration and scaling of parameters used with this method can be found in Appendix \ref{sec:config}.
Figure \ref{fig:demo1} shows the distance measure as a function of input parameters, and corresponding models for 3 quantiles.
The left and right panels present the quantile model obtained using 100 and 300 training simulations, respectively.
The colourful bands correspond to the predicted quantile functions, where band width corresponds to $\pm 1$ median absolute deviations $\sigma_d$.
That uncertainty was calculated using 128 models, each leaving out 3\% of the data at random.
The size of the uncertainty estimate $\sigma_d$ visibly decreases when the number of training samples increases.
It is clear that the accuracy of the quantile function estimation will depend on the number of training samples.

\subsection{Restricting the prior space}
\label{sec:waves_demo}

\begin{figure}
\includegraphics[width=1.0\textwidth]{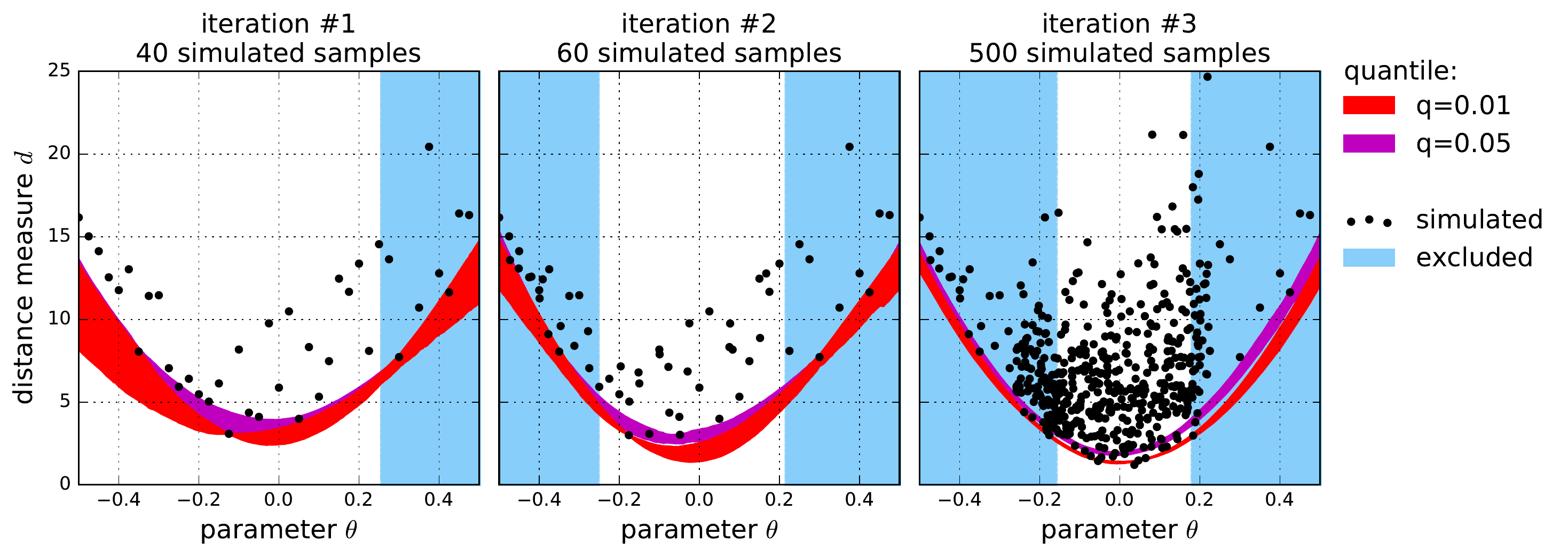}
\caption{Restricting the prior space by rejecting the infeasible regions.
Black points correspond to simulated samples.
Blue regions have been rejected using the quantile model calculated using the simulated samples on the same panel.
Left, middle and right panels correspond to iterations \#1, \#2 and \#3, which used total of 40, 60 and 500 simulated points to train the quantile model.
The allowed regions shrinks as the quantile model becomes more precise.
}
\label{fig:demo3}
\end{figure}

We now proceed to exclusion of the infeasible regions for our toy example.
We define a criterion which will determine the infeasible region: a parameter $\theta$ is excluded if
\begin{equation}
\label{eqn:feasible}
\frac{d_{q_1}(\theta) - d_{q_2}^{*}}{\sqrt{\sigma^2[d_{q_1}(\theta)] + \sigma^2[d^{*}_{q_2}]}} > n_{\sigma}
\end{equation}
where $q_1$ and $q_2$ stand for two quantiles of our choice, such that $q_1 < q_2$,
$d_{q_2}^{*}$ is the value of the $q_2$ quantile function that is the lowest among those computed for the entire prior set,
$\sigma[d_{q_1}(\theta)]$ and $\sigma[d^{*}_{q_2}]$ are the corresponding uncertainties,
and $n_{\sigma}$ is a chosen threshold level parameter.
We chose this value to be $n_{\sigma}=3$ for the rest of our analysis.
This criterion can be viewed as significance of difference between $d_{q_1}$ of the point we consider excluding and $d_{q_2}$ of the best point in the prior sample.
Our criterion can be related to the one used in \citep{wilkinson2014accelerating} in the following way.
If the noise model for the distance measure was Gaussian, then the $q_2=0.5$ quantile would correspond to the mean of the GP, and $q_1=0.0013$ to the lower $3\sigma$ GP confidence interval.
The uncertainty measured by the GP is dependent on the noise level in the data itself, as well as on the proximity of the test point to the training points; if more training points are available, the uncertainty decreases.
Quantile regression does not have this dependence, and that is why we include the additional measure of the uncertainty in our criterion; it depends on the quantile uncertainty calculated using the resampling method, described above.

\begin{table}[t]
\begin{tabular}{l}
\hline
\begin{algorithm}[H]
 \KwData{Full prior sample set from $\theta_i \sim p(\theta)$ }
 \KwResult{A classification of each sample $\theta_i$ determining whether it lies in the feasible region}
 $n_{\rm{iter}}=3$;  $n_{\rm{sim}} = [40, 20, 440]$;  $w=0$;  $\theta^w_i = \theta_i$\;
 \While{$w<n_{\rm{iter}}$}{
   select $n_{\rm{sim}}[w]$ samples $\theta^{w}_s$ from the current feasible set $\theta^{w}_i$\;
   run simulator for $\theta^{w}_s$ selected samples and calculate corresponding $d_s$\;
   train the \textsc{LiquidSVM}-QR on all [$\theta_s$, $d_s$] sets simulated so far\;
   predict $d_{q_1}(\theta_i)$, $d_{q_2}(\theta_i)$, $\sigma[d_{q_1}]$, $\sigma[d_{q_2}]$ for all prior samples $\theta_i$\;
   reject samples according to the criterion in Equation \ref{eqn:feasible}\;
   create new feasible set $\theta^{w+1}_i$ with allowed samples
  }
  evaluate simulation for all $\theta^w$\;
  set thresholds and calculate ABC posterior\;\
\caption{\abcq \ algorithm for infeasible region rejection.}
\label{alg:waves}
\end{algorithm}\\
\hline
\end{tabular}
\end{table}

Both $n_{\sigma}$ and $q_1, q_2$ control the speed of convergence of the method.
High $n_{\sigma}$ and large difference between $q_1, q_2$ will cause the method to be more conservative and will lead to slower rejection of infeasible regions.
On the other hand, low $n_{\sigma}$ and small difference between $q_1, q_2$ will cause the method to reject regions quicker, but possibly less accurately.
If these parameters are set to be too low, the algorithm may fail and cause rejection of regions which would be accepted to the posterior by the basic ABC algorithm.
If they are set to be too high, the convergence will be slow and tend towards the basic ABC formulation.
For the toy example we set $q_1=0.01$ and $q_2=0.05$.

In the toy example, we calculate infeasible regions in three iterations.
Iterations \#1, \#2 and \#3 were calculated after having simulated 40, 60 and 500 points, respectively.
Every time, all available simulations were used for training the model.
The procedure used is shown in Algorithm \ref{alg:waves}, and is inspired by the procedure in \citep{wilkinson2014accelerating}.
Figure \ref{fig:demo3} shows the distance measure as a function of the model parameter $\theta$.
The simulated points in black, the rejected infeasible regions with blue.
The quantile functions $d_q$ for $q=0.01$ and $q=0.05$ are shown with red and magenta lines, respectively.
Left, centre and right panels correspond to iterations \#1, \#2 and \#3.
The new simulations are calculated only in the regions that were designated as feasible.
The feasible region converges toward the true minimum of the function and shrinks when more simulations are included.
In this example, 24\% of the prior volume was excluded after iteration \#1, 51\% after \#2 and 64\% after \#3.

\subsection{Using single-dimensional projections}
\label{sec:projections}

Our toy model used a single-dimensional input parameter $\theta$.
For most problems, however, the dimensionality of $\theta$ will be higher, and the quantile model will be created in this high-dimensional space.
In such case, the high-dimensional correlations between the distance measure values will be exploited.
The drawback of high-dimensional model is that it needs more training data to compute accurate quantiles.
This is an example of the ``curse of dimensionality''.
On the other hand, even a single-dimensional marginal can be used for rejecting infeasible regions.
It is possible to apply Algorithm \ref{alg:waves} to a set of lower-dimensional marginals of $p(d|\theta)$.
The advantage of using a single dimension is that a precise quantile model can be trained with comparatively fewer training points.
A successful strategy will therefore use single-dimensional marginals early on, to exclude infeasible regions in 1D as fast as possible.
When more simulations are available, the full dimensional input can be used to exploit high-dimensional correlations inside $p(d|\theta)$.
Given that training of \textsc{LiquidSVM} models is usually very fast compared to the evaluation of simulations, even high number of combinations of parameter marginals can be used.

\section{Application to the estimation of redshift distributions}
\label{sec:application}

We proceed to presenting the application of this method estimation of redshift distributions of cosmological samples from \citetalias{herbel2017redshift}.
Our goal was to apply the \abcq\ algorithm to find out if it can give the same solution as the basic ABC method, and possibly do it with fever simulations.
We did not run any additional simulations in this work and used only those created in \citetalias{herbel2017redshift}.
Henceforth throughout this work we will refer to ``running a simulation'' as though it was a real simulator, although in reality we use another data point from the results of \citetalias{herbel2017redshift}.
We always keep track of how many points were ``simulated'' this way, as it is our objective to estimate the potential improvement in speed achieved by our method.

\subsection{Data description}
\label{sec:data_description}

In this section we briefly summarise the data used in \citetalias{herbel2017redshift}.
The aim of that work was to measure the distribution of redshifts of samples of galaxies detected in the COSMOS field \citep{capak2007first,taniguchi2006cosmic}.
The image data was obtained from the publicly available image data from the Suprime-Cam imager on the Subaru telescope.
Additionally, both wide and deep spectroscopic catalogue of galaxies from the VVDS Survey\footnote{\url{http://cesam.lam.fr/vvds/}} was used \citep{lafevre2013vimos}.
Magnitudes and sizes of galaxies were measured from the image data using the \textsc{SExtractor} code \citep{berin1996sextractor}.

The imaging data was simulated with the Ultra Fast Image Generator (\textsc{UFig}) \citep{berge2012ultra}.
The parametric model included:
(i) luminosity functions of galaxies, which describe the distribution of number counts of galaxies with particular brightness at given redshift,
(ii) the distribution of galaxy size as a function of brightness, and
(iii) parameters describing the relation between the spectra of galaxies.
The priors on these parameters were taken from previous measurements using large galaxy samples in \citep{beare2015optical,mandelbaum2014third}, and were broadened to allow more freedom to the ABC algorithm to explore this parameter space.
The number of simulated samples was 140000 and spanned 31 parameters.
In this work, we neglect the parameters controlling the colours of objects, and keep 11 parameters corresponding to luminosity functions and size/luminosity dependencies.
The reason is that we have little constraining power on the colour parameters, and effectively marginalise these parameters out in the ABC process.
In other words, they can be considered as another source of noise in the problem.

The final quantity of interest is the distribution $n(z)$ of redshifts $z$ of galaxies observed in the simulated image data.
This distribution is used to understand the sample of observed galaxies and estimate expected cosmological signals, such as, for example, weak gravitational lensing shear \citep{hildebrandt2012cfhtlens,bonnett2016redshift}.
Images are simulated using models characterised by input parameter $\theta$; for each $\theta$ we obtain a $n(z)$ distribution.
A posterior distribution on $\theta$ will therefore give a family of $n(z)$ distributions and thus give an uncertainty on the measured $n(z)$.

Five distance measures were calculated in \citetalias[see][for details]{herbel2017redshift}, based on the number counts of detected galaxies and the distributions of their properties, such as brightness, size, colours and redshift.
In that work, the set of five thresholds was calculated using the following algorithm:
a threshold corresponded to a quantile $Q$, such that there were 150 samples satisfying the condition $d^j_i<d_Q^{j}$, where $d_Q^{j}$ is the value of the distance measure corresponding to quantile $Q$, $i$ is the sample index and $j$ is the distance measure index.
The quantile $Q$ was calculated using all 140000 simulated samples.
Here, we use an alternative procedure for combining distances: we create a single distance measure that combines all five.
First, to bring all distance measures to the same numerical range, we divide them by a factor $d^{j}_{10}$, which corresponds to $10$-th percentile found using first 500 samples we simulated in our process.
The rescaled distance $d^{j}_{s}$ is simply $d^{j}_{s} = d^{j} / d^{j}_{10}$, where $d^{j}$ is the original distance, as measured by \citetalias{herbel2017redshift}.
To create the combined distance $d_{c}$, we apply the operation $d_{c}=\max_j d^{j}_{s}$.
This single distance measure is used to create the posterior distribution.
We found, that the results were not sensitive to the choice of the operation used to combine distances (maximum in our case).
For example, using a mean operation gave a very similar posterior.
Also, the choice of 10\% percentile did not play a significant role within that method, with the posterior mean redshift changing slightly when different values were used.
We also found that using this distance combination procedure results in a slightly different mean redshift $\hat z$ than reported in \citetalias{herbel2017redshift}.
This difference was $\sim 0.3 \sigma$, and the standard deviation on $\hat z$ also changed by $\sim 15\%$  (see Section \ref{sec:results}).
This is not unexpected, since a change of distance combination method has an effect on the relative weighting of distances.
We did not investigate these differences as both methods produce statistically consistent results.

\subsection{Application}

We create the following training strategy for the redshift distribution measurement problem.
We use 12 \abcq\ models jointly: 11 single dimensional ones, and a single 11-dimensional one.
A point in parameter space is rejected if at least one of the models decides to do so.
We did not explore more complicated application strategies, leaving it to future work.
We used the \abcq\ model with quantiles $q_1=0.01$ and $q_2=0.5$ (see Section \ref{sec:method} and Appendix \ref{sec:config}) for 11-dimensional model, and $q_1=0.01$ and $q_2=0.05$ for single dimensional projections.
We trained the quantile model in 15 iterations.
First and second iterations used 500 and 1000 samples, respectively, and each following iteration was calculated after 2000 new simulations were available.
After 15 iterations the number of points in feasible region was 25235.
The posterior was then calculated using 150 best samples according to the combined distance measure (see Section \ref{sec:application}).

\subsection{Results}
\label{sec:results}

We compare the \abcq \ method to the basic ABC algorithm.
The performance is considered to be better if an algorithm converges to the final posterior distribution using fewer simulation runs.
Instead of comparing the width of the 11-dimensional posterior on $\theta$, we focus on the uncertainty on the $n(z)$ for this posterior (see Section \ref{sec:data_description}).
For each sample from the posterior, we calculate the mean redshift $\hat z = \langle n(z) \rangle$.
We characterise the redshift uncertainty of the posterior using a single quantity: the standard deviation $\sigma[\hat z]$ of the means of the redshift distributions.
We calculate $\sigma[\hat z]$ as a function of the number of simulations ran for both \abcq \ and the basic method.

\begin{figure}\centering
\includegraphics[width=0.6\textwidth]{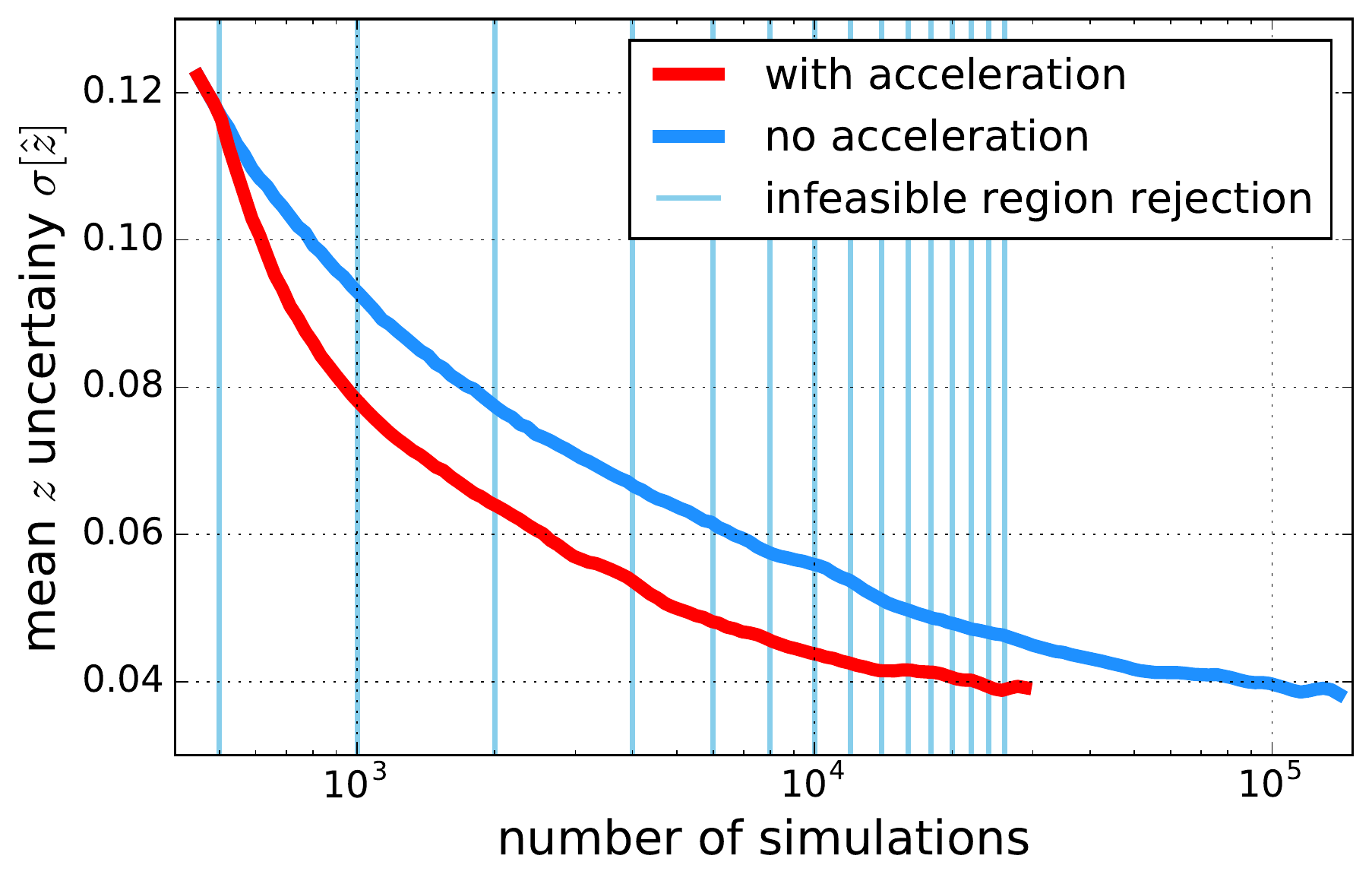}
\caption{Convergence of the \abcq\ algorithm compared to the basic ABC.
The blue line shows the uncertainty on the mean redshift $\sigma(\hat z)$ as a function of number of simulated samples from the prior, for the basic ABC algorithm.
The red line shows the uncertainty on $\hat z$ when \abcq\ algorithm is used.
The model was calculated 15 times.
The moments when the model was calculated are marked with light blue vertical lines.
In this example, the \abcq\ method needed only $\sim$29500 samples to achieve the same uncertainty on $n(z)$ as the basic algorithm with 140000 simulated samples.
}
\label{fig:convergence}
\end{figure}

Figure \ref{fig:convergence} presents the convergence of the algorithms in terms of $\sigma[\hat z]$ of the posterior as a function of the number of evaluated simulations.
We randomised the calculation of $\sigma[\hat z]$ over the order in which the simulations may be obtained; for every number of simulations we took a median $\sigma[\hat z]$ out of 5000 random permutations of the order.
The convergence of the basic ABC algorithm is shown in the thick blue line and that of the \abcq\ method in red.
The vertical light blue lines show the moments when the model was trained and applied to reject parts of prior space.

\begin{figure}\centering
\includegraphics[width=0.75\textwidth]{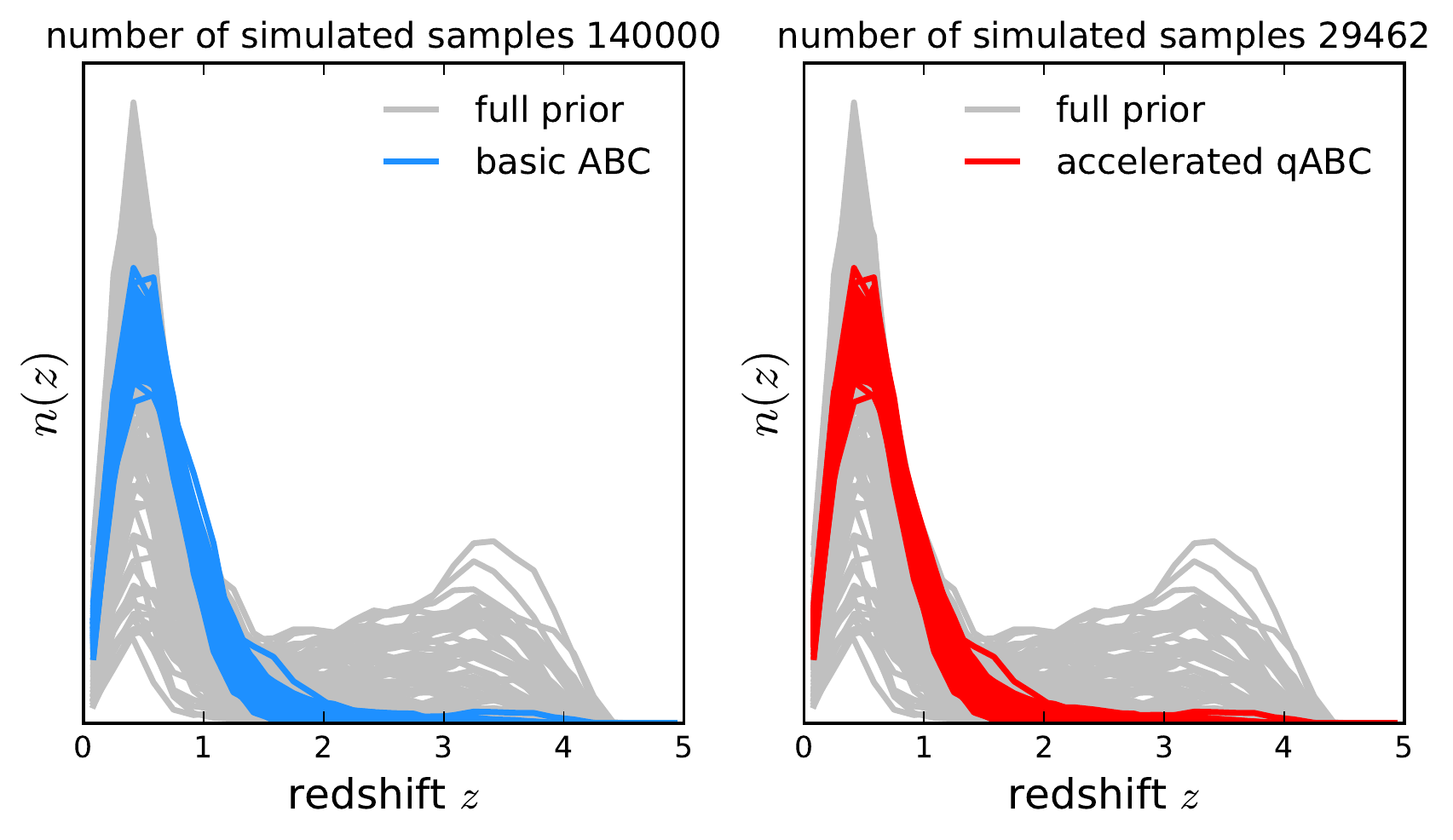}
\caption{Comparison between the posterior $n(z)$ calculated using the basic ABC method (left) and accelerated, \abcq\ method (right).}
\label{fig:comparison}
\end{figure}

The \abcq\ method obtains $\hat z=0.633 \pm 0.039$, whereas the basic ABC obtains $\hat z=0.633 \pm 0.038$; the results are almost identical.
Figure \ref{fig:comparison} shows the $n(z)$ distributions for the ABC posterior for the basic and the accelerated algorithm.
These posteriors differ by 3 out of 150 samples only (2\%).
These 3 samples are incorrectly classified as infeasible.
This may be due to small modelling errors in quantile regression and is not unexpected.
Given that this fraction is very small, we do not investigate this further; the precision can always be increased by using more conservative settings (see Section \ref{sec:toy}).

The accelerated algorithm used 29462 simulator runs: 25235 inside the final feasible region, plus 4227 outside, which were used for training at earlier iterations.
This is $\sim$20\% of the 140000 simulations used by the basic ABC, which constitutes almost fivefold acceleration.
This acceleration rate is specific for this particular problem with our choice of priors.
Specifically, if a wider prior was used, we would expect a higher acceleration rate.
For other prior choices and for other problems, the acceleration rate may vary.

\begin{figure}\centering
\includegraphics[width=1\textwidth]{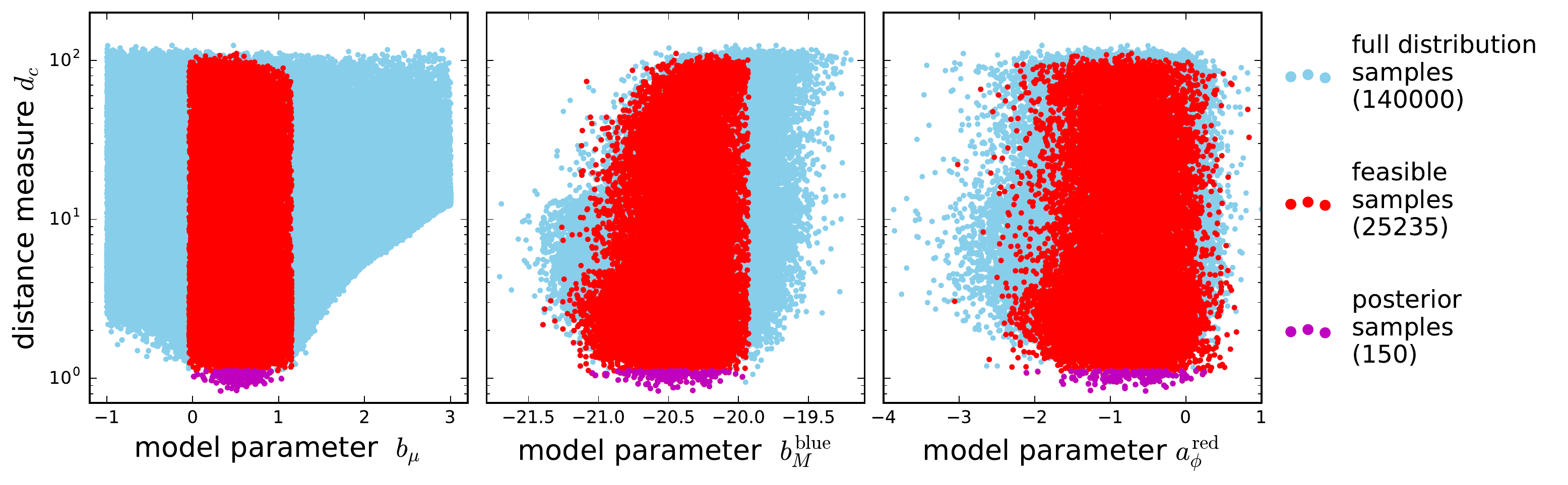}
\caption{Distance measure distribution as a function of model parameters for:
(left) the intercept of the galaxy size-magnitude relation $b_{\mu}$,
(middle) the intercept of the redshift dependence of $M_{*}$ component of the luminosity function for blue galaxies $b_{M}^{\rm{blue}}$,
and (right) the exponential decay rate with redshift of the $\phi_{*}$ parameter of the luminosity function for red galaxies $a_{\phi}^{\rm{red}}$.
Light blue points represent the distribution found by \citetalias{herbel2017redshift}: 140000 samples with distance measures.
The red points show 25235 feasible samples as determined by the final \abcq\ model, which was trained using 26000 samples.
The algorithm rejected regions which have very low probability of being accepted into the posterior distribution.}
\label{fig:quantiles_1d}
\end{figure}

To illustrate how the \abcq\ algorithm works for this application, we consider the distribution of the distance measure as a function of model parameters, before and after infeasible regions rejection.
Figure \ref{fig:quantiles_1d} shows the distance measure for 3 parameters:
the intercept of the galaxy size-magnitude relation $b_{\mu}$ (left panel),
the intercept of the redshift dependence of $M_{*}$ component of the luminosity function for blue galaxies $b_{M}^{\rm{blue}}$ (middle panel),
and exponential decay rate with redshift of the $\phi_{*}$ parameter of the luminosity function for red galaxies $a_{\phi}^{\rm{red}}$ (right panel; see equations 3.1-3.4 in \citetalias{herbel2017redshift} for details about model parameters).
The full distribution from \citetalias{herbel2017redshift}, before rejection, is shown in blue points.
The final feasible region, after rejection, is shown in red points.
The samples accepted into the posterior are shown in magenta points.
We notice that the algorithm indeed rejects regions that have very low chance of being accepted into the posterior, and allows further sampling only in regions which contain the minimum of the distance measure function.
The feasible regions are just a little broader then the final posterior, which demonstrates the effectiveness of our approach.
These plots show one dimension at a time, with the remaining 10 dimensions marginalised.

\section{Conclusions}
\label{sec:conclusions}
We have proposed a novel method to accelerate Approximate Bayesian Computation with the use of Quantile Regression, which we call \abcq.
This method aims to model the distribution $p(d|\theta)$ of the distance measure $d$ given a set of model parameters $\theta$, with the assumption that this distribution varies smoothly with $\theta$.
This model can be created with a relatively small number of training samples.
The model of quantiles of $p(d|\theta)$ is calculated in high dimensional $\theta$ parameter space with the use of the Support Vector Machine implementation in \textsc{LiquidSVM}.
Once the model is created, it can be used to quickly reject parts of the prior space, for which the distance measures are large, and that would not be accepted to the posterior.
This way, we avoid running costly simulations in infeasible regions of the prior, and thus save computing time and accelerate the ABC method.

We applied this method to the problem of the measurement of the redshift distribution of cosmological samples, as presented in \citetalias{herbel2017redshift}.
That work used a basic implementation of the ABC algorithm, without using any acceleration technique.
The number of samples, for which the simulator was run, was 140000.
We used this dataset to find out whether the \abcq\ method could converge to the same posterior distribution as the basic ABC method, with the use of fewer simulations.
After each iteration, the prior space was restricted and a part of parameter space was rejected by the model.
These regions of parameter space were then excluded from further analysis.
We found that after 15 iterations the number of feasible samples was approximately 25000.
If the simulations were ran for these feasible samples, the posterior on $n(z)$ would be almost the same as that calculated with the basic ABC method, after having simulated 140000 samples.
Adding together feasible samples and those used for training, the total is roughly 29500 simulations.
That indicates that, in our measurement, the \abcq\ method was able to converge using only $\sim$ 20\% of the samples needed by the basic ABC for the same accuracy.
The final posterior obtained from the accelerated ABC and the basic ABC methods are almost the same.
This result constitutes a significant improvement in the ABC speed.
For other problems and different prior configurations the level of acceleration may vary.
The acceleration may also vary with the settings of the \abcq\ algorithm and SVM engine (see Section \ref{sec:toy} and Appendix \ref{sec:config}).

Another advantage of the \abcq\ method is that it is trivially parallelisable.
This can be important for ABC, as it is often ran with large simulations which use codes with their own parallelisation schemes.
Some MCMC-based ABC methods are thus hard to parallelise, as the next sample is chosen solely based on previous one.
In \abcq\, the simulations can be evaluated completely independently, and the quantile model can be applied only from time to time, as the number of available simulations increases.

Several future improvements to this method are worth exploring.
It may be possible to extend the quantile regression model to include the uncertainty estimate on the quantile function.
The algorithm presented in this work creates an uncertainty estimate by a simple resampling method: quantile regression is run many times, each time omitting 3\% of the available training data.
A probabilistic quantile regression method, perhaps similar in nature to Gaussian Processes, could estimate these uncertainties more naturally.

Finally, the posterior could be estimated by directly sampling from the model of $p(d|\theta)$, instead of using the model only to restrict the parameter space.
A dense grid of quantiles could be used to estimate the full cumulative probability $P(d|\theta)$.
Once the model has been constructed, for some value of a chosen threshold, the probability density of the posterior for $\theta$ will be proportional to the quantile of $P(d|\theta)$ at that threshold.
Samples can be drawn from that probability.
Using such a scheme can potentially allow to reach lower thresholds faster than with the current method.

\acknowledgments

We would like to thank Philipp Thomann for help with the \textsc{LiquidSVM} package.
Special thanks to Amirreza Bahreini and Armin Van de Venn for work on related problem.
This work was supported in part by grant number $\rm{200021\_169130}$ from the Swiss National Science Foundation.

\bibliography{refs}{}
\bibliographystyle{unsrt}

\appendix
\section{SVM usage and configuration}
\label{sec:config}
As kernel methods are not scale invariant, it is customary to scale the data to a desired numerical range.
When using SVM, we always pre-process the input parameters, such that the minimum of each parameter is 0 and maximum 1.
For distance measures, we apply transformation $y_{\rm{transformed}}=y(10+1)/(y+10)$.
This guarantees that the numerical range for distance measures to be roughly between 0 and 100.
This function has a minimum of 0, is almost linear for small distances and behaves similarly to a logarithm for very large distances.
Such functional form provides stability to the SVM algorithm.
After the SVM run, the parameters are transformed back.
Throughout this work we run \textsc{LiquidSVM} with the following settings:
\verb+scale=False, grid_choice=1+, \verb+adaptivity_control=0+, \verb+useCells=False+, \verb+retrain_method=select_on_entire_train_set+.

\end{document}